\documentclass[twocolumn,prb,showpacs,floatfix,superscriptaddress]{revtex4}%
\usepackage{graphicx}
\usepackage{dcolumn}
\usepackage{bm}
\usepackage{amssymb}
\usepackage{amsmath}
\usepackage{amsfonts}
\usepackage{xcolor}%
\providecommand{\U}[1]{\protect\rule{.1in}{.1in}}
\newcommand{\nalio}{(Na$_{1-x}$Li$_x$)$_2$IrO$_3$}
\newcommand{\liiro}{Li$_2$IrO$_3$}
\newcommand{\nairo}{Na$_2$IrO$_3$}
\providecommand{\U}[1]{\protect\rule{.1in}{.1in}}
\begin{document}
\title{Effect of isoelectronic doping on honeycomb lattice iridate A$_{2}$IrO$_{3}$}
\author{S. Manni}
\affiliation{I. Physikalisches Institut, Georg-August-Universit\"at G\"ottingen,D-37077,
G\"ottingen, Germany }
\author{Sungkyun Choi}
\affiliation{Clarendon Laboratory, University of Oxford, Parks
Road, Oxford OX1 3PU, United Kingdom}
\author{I. I. Mazin}
\affiliation{Code 6393, Naval Research Laboratory, Washington, DC 20375, USA}
\author{R. Coldea}
\affiliation{Clarendon Laboratory, University of Oxford, Parks
Road, Oxford OX1 3PU, United Kingdom}
\author{Michaela Altmeyer}
\affiliation{Institut f\"ur Theoretische Physik, Goethe-Universit\"at Frankfurt, 60438
Frankfurt am Main, Germany}
\author{Harald O. Jeschke}
\affiliation{Institut f\"ur Theoretische Physik, Goethe-Universit\"at Frankfurt, 60438
Frankfurt am Main, Germany}
\author{Roser Valenti}
\affiliation{Institut f\"ur Theoretische Physik, Goethe-Universit\"at Frankfurt, 60438
Frankfurt am Main, Germany}
\author{P. Gegenwart}
\affiliation{I. Physikalisches Institut, Georg-August-Universit\"at G\"ottingen,D-37077,
G\"ottingen, Germany }
\date{\today}

\begin{abstract}
We have investigated experimentally and theoretically the series
(Na$_{1-x}$Li$_{x}$)$_{2}$IrO$_{3}$. Contrary to what has been
believed so far, only for $x\leq0.25$ the system forms uniform
solid solutions where Li preferentially goes to the Ir$_{2}$Na
planes as observed in our density functional theory calculations
and confirmed by X-ray diffraction analysis. 
For larger Li content, as evidenced by powder
X-ray diffraction, scanning electron microscopy and density
functional theory calculations, the system shows a miscibility gap
and a phase separation into an ordered Na$_{3}$LiIr$_2$O$_{6}$
phase with alternating Na$_3$ and LiIr$_2$O$_6$ planes, and a Li-rich
phase close to pure Li$_{2}$IrO$_{3}$. For $x\leq 0.25$  we
observe (1) an increase of $c/a$ with Li doping up to $x=0.25$,
despite the fact that $c/a$ in pure Li$_{2}$IrO$_{3}$ is smaller than in
Na$_{2}$IrO$_{3}$, and (2) a gradual reduction of the
antiferromagnetic ordering temperature $T_{N}$ and ordered moment.
The nature of the previously proposed  continuous magnetic quantum phase transition (QPT) at
$x\approx 0.7$ needs to be re-evaluated as the proof of miscibility gap in \nalio~phase diagram is inevitable.
\end{abstract}

\maketitle

\hspace{5.2in}

\section{Introduction}
Quasi-2D correlated oxides with honeycomb
layers have been attracting considerable interest in the last
years~\cite{Singh10,Choi2012} largely because of their capacity to
host interesting topological and frustration
phenomena~\cite{Shitade2009,Jackeli1}.  Of particular interest is
Na$_{2}$IrO$_{3},$ where several critical energy scales are
comparable, such as one-electron hopping $t,$ Hubbard repulsion
$U,$ Hund's rule coupling $J,$ and spin-orbit interaction
$\lambda.$ A possible, albeit not necessary, consequence of the
competition between several comparable energy scales is strong
frustration, in particular magnetic, which may lead to long-sought
spin-disordered phases at zero temperature.

It was recently proposed~\cite{Cao} that Na$_{2}$IrO$_{3}$ and
Li$_{2}$IrO$_{3}$ form a continuous solid solution, with the
N\'{e}el ordering temperature maximized in the end compounds and
going near to zero at an intermediate doping,
(Na$_{1-x}$Li$_{x})_{2}$IrO$_{3},$ $x\sim0.7.$ Such a  quantum
phase transition would be of great interest, as it would allow
going from a quantum spin liquid state to different types of long
range order by changing doping in two different directions.

In this work we show, both experimentally and theoretically, that
the assumption of a continuous solid solution is not justified. In
particular, for $x>0.25$ the system experiences a phase
separation, which has a profound physical reason. Specifically we
find that the $x=0.25$ state, namely the one where all Na in the
Ir$_{2}$Na planes are substituted by Li while  Na$_{3}$
plane remains intact, is exceptionally stable.

This stability is gained through the fact that Li is smaller than
Na and therefore allows shorter Ir-Ir bond lengths, when placed in
the same plane. Indeed, as was observed
earlier,~\cite{Jackeli1,Mazin12,Foyevtsova2013} two different
Ir-Ir hoppings compete in this system: direct overlap of the like
orbitals, and indirect, O-assisted hopping of unlike orbitals.
Even small changes in geometry affect this competition
dramatically. On the other hand, partial substitution of the
interlayer Na by Li is not energetically favorable because the
interlayer separation is defined by the larger Na ions and is not
optimal from the Li point of view. This is why compositions with
$x>0.25$ prefer phase separation.

We also observe a N\'{e}el temperature reduction with increasing
doping up to $x<0.25$ as was previously reported.~\cite{Cao} In
fact, our findings on the underlying doped lattices are essential
to understand both the N\'{e}el temperature reductions as pure
end-members are respectively doped (it is likely that the
mechanisms are different for the Na-rich and Li-rich alloys), and
the nature of the putative quantum critical point~\cite{Cao}. Most importantly, we observe a chemical phase separated region in the \nalio~phase
 diagram for $x>0.25$ (extending to at least $x=0.6$), which question a continuous QPT at $x=0.7$ as suggested Cao et. al.~\cite{Cao}.

\section{Experimental details} 
Single crystals of {(Na$_{1-x}$Li$_{x}$)$_{2}$IrO$_{3}$} have been grown using a
similar procedure as previously used for
{Na$_{2}$IrO$_{3}$}~\cite{Singh10}. A first calcination process
has been done at 750$^{\circ}$C with stoichiometric proportions
of carbonates (Na$_2$CO$_3$ and Li$_2$CO$_3$) and Ir metal. After
prereaction at 900$^{\circ}$C the polycrystalline material was
processed for crystal growth with excess IrO$_{2}$ flux. The
amount of excess IrO$_{2}$ and the temperature of crystal growth
were varied for different doping levels. Since with increasing Li
content the solubility of  the phase in the flux decreases, it is
important to control both temperature and excess IrO$_2$ for
obtaining large enough crystals for bulk measurements.

\begin{table}[h]
\caption{Comparison between the nominal and actual Li content determined by
ICPMS in \% of Li in (Na$_{1-x}$Li$_{x}$)$_{2}$IrO$_{3}$}%
\label{table:ICPMS}%
\centering
\begin{tabular}
[c]{ccc}\hline\hline
x & Nominal Li (\%) & ICPMS Li (\%)\\[0.5ex]\hline
0.05 & 5 & 3.83 ($\pm$0.2)\\
0.1 & 10 & 9.5 ($\pm$0.5)\\
0.2 & 20 & 21.8 ($\pm$1.5)\\
0.3 & 30 & 33.2 ($\pm$1.1)\\
0.4 & 40 & 47.0 ($\pm$0.9)\\[1ex]\hline
\end{tabular}
\end{table}

The Na:Li ratio was determined by inductively coupled plasma mass
spectrometry (ICPMS) on different pieces of crystals of every doping
level. In contrast to the claim of
Ref.~\onlinecite{Cao} we have found that it is not possible to detect Li by an energy dispersive X-ray
(EDX) analysis since Li is a light metal. In EDX we can only observe changes in the Na to
Ir ratio, which decreases with Li doping. Table~\ref{table:ICPMS}
gives a comparison between the nominal (starting composition) and the
measured Li fractions.  Some of the plate-like crystals were crushed
and powder x-ray diffraction (XRD) was performed for the scattering
angle range $10^{\circ} \leq 2\theta \leq 100^{\circ}$ with Cu
K$_\alpha$ radiation to estimate the change of the lattice parameters
with Li doping. Single crystal x-ray diffraction (XRD) was performed
using a Mo-source Oxford Diffraction Supernova diffractometer on
crystals of (Na$_{1-x}$Li$_{x}$)$_{2}$IrO$_{3}$ with nominal doping
$x$ from 0.05 to 0.4 in order to obtain lattice parameters and confirm
the crystal structure and internal atomic coordinates. The samples
were thin, plate-like crystals with a typical size of $
70\times60\times10~\mu$m$^{3}$.  Magnetization, ac susceptibility and
specific heat were measured in commercial SQUID magnetometer and
physical property measurements systems, respectively.

Since the size of Li-doped crystals decreases with doping, we have
used lumps of crystals for magnetization and specific heat
measurements. Crystals (or lumps) have been separated
mechanically. Sometimes some remaining flux is present in the lump
which gives a low temperature Curie tail in the $\chi(T)$
measurement.

\section{Theoretical calculations \label{theo}} 
In order to determine the most realistic doped structures, we
performed structural relaxations on supercells of
{(Na$_{1-x}$Li$_{x}$)$_{2}$IrO$_{3}$} for Li dopings $0 \leq x \leq 1$
in steps of $0.125$ within density functional theory (DFT).  We
considered the 
generalized gradient approximation (GGA) as exchange-correlation 
functional and employed the projector augmented wave
(PAW) basis set as implemented in the Vienna ab initio simulation
package.~\cite{VASP} An $8\times6\times8$ $k$ mesh was used.
Since (i) the end
compounds Na$_2$IrO$_3$ and Li$_2$IrO$_3$ show long range magnetic order
and (ii) Ir is a $5d$ ion, magnetism,
correlation and spin-orbit coupling (SOC) effects  may be
important for precise structure predictions.
  However, DFT calculations including spin-orbit coupling
 are very time consuming.  We followed therefore the following
strategy. For the rather expensive determination
of the most stable configurations for each doping level
  we initially considered
the GGA functional without inclusion of SOC and
magnetism.  The information gained from these
results was subsequently used  to perform more elaborate
calculations including spin-orbit coupling, a Hubbard repulsion
$U=3$~eV and spin polarization (spin-polarized GGA+SOC+U). We found
that while these more precise calculations lead to much better
comparison of lattice parameters with experiment, at the qualitative
level the plain GGA calculations seem to be sufficient.

 For our GGA calculations we considered all possible Li configurations in a
unit cell containing four formula units and searched for the most
stable case. In order to verify the stability of the configurations,
we also considered for some dopings supercells of sizes
2$\times$1$\times$1 and 1$\times$2$\times$1 where the unit cell with
four formula units was doubled along $a$ and along $b$ respectively.
The total energy calculations obtained with the PAW basis
were double-checked against the all
electron full potential local orbital (FPLO) code~\cite{FPLO} (see
 Fig.~\ref{phasewithoutSOC} in Appendix B).

\begin{figure}[ptb]
\centering
\includegraphics[width=0.8\columnwidth]{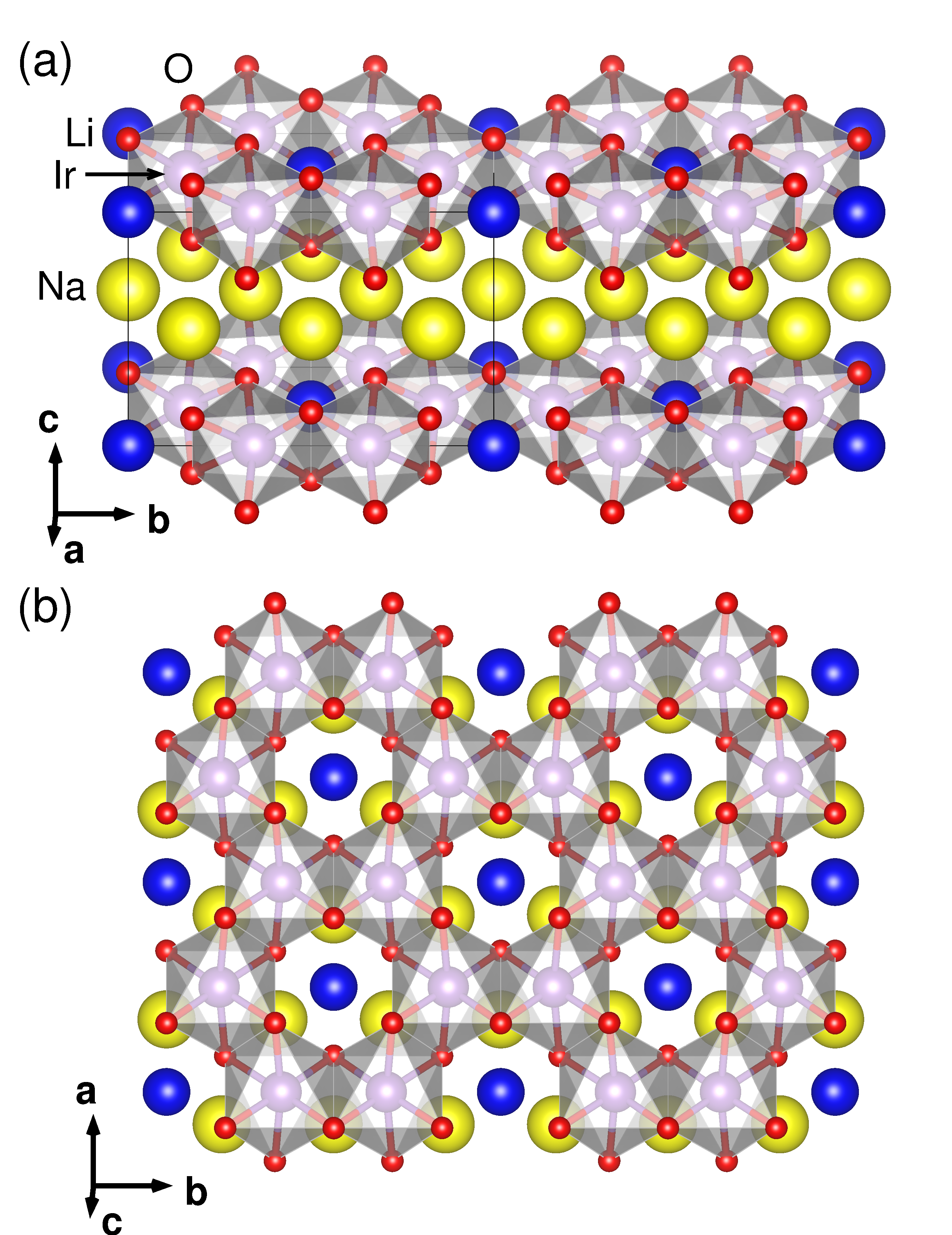}
\caption{(Color online) Calculated crystal structure of
(Na$_{1-x}$Li$_{x}$)$_{2}$IrO$_{3}$ for $x=0.25$: (a) layered structure of Ir$_2$Li and Na$_3$ planes and (b) view on the Ir$_2$Li planes, where the Ir atoms form a honeycomb lattice} \label{structures}
\end{figure}

In our search for optimally relaxed structures,  we considered two types of
calculations. In one set of calculations the lattice parameters were
fixed to the experimentally determined values (see Fig.~\ref{XRD}(c))
and the internal coordinates were relaxed. In the second set of
calculations we performed a full relaxation including both volume and
internal coordinates. Both calculations showed that for $0 \leq x \leq
0.25$ the energetically most favorable location for Li ions are Na
positions in the honeycomb layer. 
In Fig.~\ref{structures} we present the most stable crystal
structure of (Na$_{1-x}$Li$_{x}$)$_{2}$IrO$_{3}$ for
 a doping level of $x=0.25$. For both types of relaxations (at fixed volume
and including volume relaxation)
the highest stability was obtained for Li
substituting Na in the Ir$_2$Na planes rather than in the Na$_3$
planes.
Further doping leads to a replacement of Na atoms in the Na$_3$ layer,
 where we found clustering of the Li atoms to be energetically favorable.
 This observation is also supported by the consideration
 of  supercells containing eight formula units at a doping level of $x=0.5$.
In this case we found the structures with most
clustering to be lowest in energy, while the configurations with
a homogeneous distribution of Li atoms in the Na$_3$ layer are
about 50~meV/f.u. (within GGA) higher in energy compared to the configurations
with clustering.\\

\section{Results and Discussion}
\subsection{Low doping ($x<0.25$)}

\begin{figure}[h]
\centering
\includegraphics[width=\columnwidth]{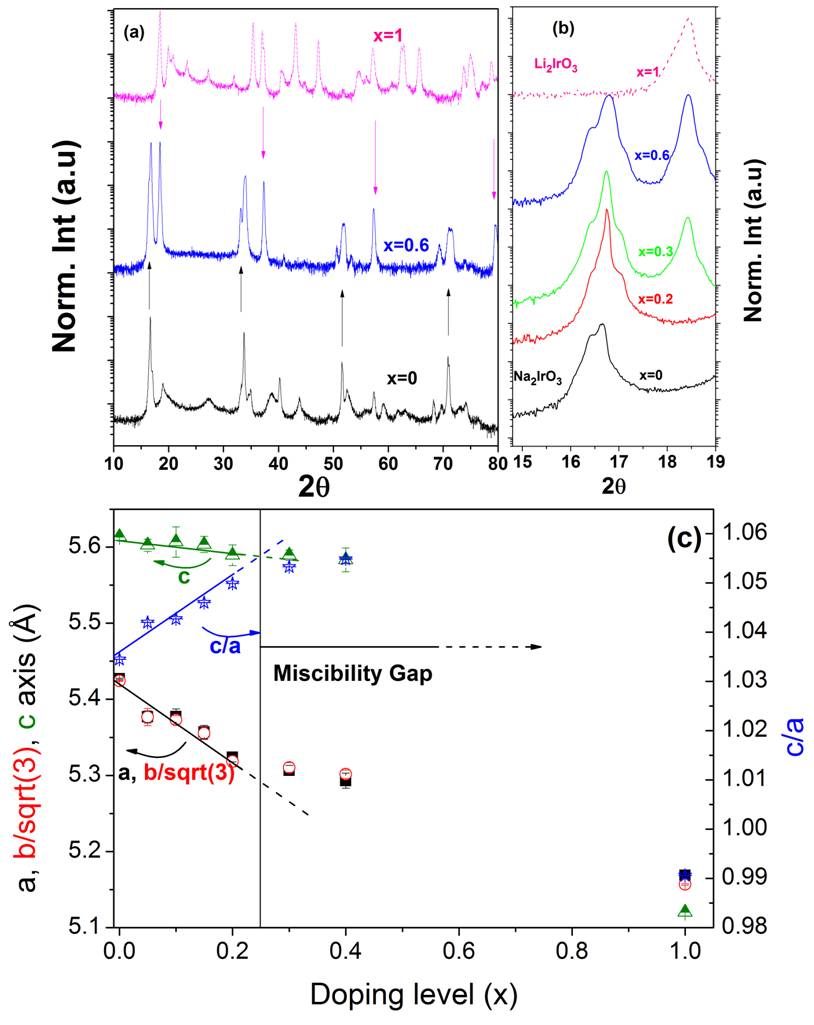}
\caption{(Color online) (a) Powder XRD of the crushed
(Na$_{1-x}$Li$_{x}$)$_{2}$IrO$_{3}$~crystals for x=0, 0.6 and x=1. The magenta colored downward arrows point x=0.6 XRD peaks that matches with x=1 (00n) peaks and black colored upward arrows point x=0.6 XRD peaks that matches with x=0 (00n) peaks. (b) Zoomed XRD spectra in the 2$\theta$ region 15 to 19 $^{\circ}$ for all values of x. (c) Lattice
parameters obtained from single crystal XRD of
(Na$_{1-x}$Li$_{x}$)$_{2}$IrO$_{3}$ single crystals ($x=1$
obtained from Ref.~\onlinecite{Singh12}). The horizontal arrow
marks the miscibility gap region where samples showed phase
separation. Solid straight lines (extended by dashed lines in the
miscibility gap region) are guides to the eye.}. \label{XRD}
\end{figure}

\textbf{Structural changes:} Powder XRD of crushed
(Na$_{1-x}$Li$_{x}$)$_{2}$IrO$_{3}$ crystals shows single phase
crystals up to $x=0.2$  (see Fig.~\ref{XRD}(b)). These crystals
are very plate-like and only ($00n$) peaks could be observed.
Moreover, while ICPMS confirms the inclusion of Li (see
Table~\ref{table:ICPMS}) at the concentration $x=0.2$, there is
almost no shift of the (001) peak, implying almost no change in
the $c$ lattice parameter for the range $0 \leq x \leq 0.2$.

The lattice parameters as a function of doping were determined by
single crystal XRD. Complete diffraction patterns
for structural refinement were collected for the best samples at
each doping. We faced nevertheless a few challenges
when  refining the  diffraction pattern of the Li-doped samples.
 Namely, Li scatters x-rays very weakly and its precise position in the
structure cannot be uniquely determined  from x-ray
measurements alone, especially at low Li concentrations and in the
presence of dominant scatterers like Ir (with 77 electrons),
refinements of the crystal structure with Li in different Na
positions (in the honeycomb Ir$_2$Na layer and in the hexagonal
Na$_3$ layer) gave rather similar results. Since structural
relaxation calculations (see previous section) suggest a
strong energetic preference for the doped Li to replace the Na in
the Ir honeycomb layers (for $x\leq 0.25$), the final structural
refinement (within Sir-92 and Shelx packages~\cite{single_ref})
was performed assuming that Li randomly replaces Na at this site.
The refinement converged well only when some finite degree of site
mixing ($f>0$) was assumed also on the nominally Ir honeycomb
site, so that the occupation at this site was assumed to be
$(1-f)$Ir+$f$Na. In order to preserve the total atomic
count the honeycomb center site occupation was assumed to be
$4x$Li$+(1-4x-2f)$Na$+2f$Ir. The refined atomic positions are
listed in Tables~\ref{tab_NaLi213_5per} to~\ref{tab_NaLi213_20per}
for the doping concentrations $x=0.05$ to 0.2 (Appendix A).

In order to determine
the lattice parameters accurately we measured for
each doping between 10 to 20 samples 
 and the obtained average values  are plotted in
Fig.~\ref{XRD}(c) with the error bars indicating the spread of
values for each nominal composition.  Throughout the range
 $0.05 \leq x\leq 0.2$, the
diffraction patterns show sharp peaks that could be well
indexed and refined with a $C2/m$ crystal structure derived from
the undoped ($x=0$) parent Na$_2$IrO$_3$ in
Ref.~\onlinecite{Choi2012}. For lower dopings $x=0.05,0.1$
we found samples  where the diffraction patterns could be
consistently indexed in terms of a single crystal (no twins).  For
dopings
$x=0.15,0.2$, samples showed two or three co-existing twins and in
this case refinement was successfully performed using multi-twin
techniques with the same unit cell parameters and crystal
structure for all co-existing twins. Throughout the range $0.05 \leq x \leq 0.2$
the $C2/m$ crystal structure of parent Na$_2$IrO$_3$ provides
a good description of the observed diffraction pattern, confirming
single-phase crystals with this structure. Both the $a$ and $b$
lattice parameters strongly decrease at the same rate with
increasing doping ($b/\sqrt{3}\simeq a$, which confirms a globally
almost undistorted honeycomb Ir structure in the low Li doped
region) while the $c$ parameter remains almost constant
(Fig.~\ref{XRD}(c)). Remarkably, the $c/a$ ratio increases with
increasing doping $x$ up to 0.2 (Fig.~\ref{XRD}(c)) while it is
reduced by 5$\%$ in fully-doped ($x=1$) Li$_{2}$IrO$_{3}$ compared
to the undoped ($x=0$) Na$_{2}$IrO$_{3}$. We conclude that there is no
effective $c$-axis pressure in the low Li doping region.

\begin{figure}[ptb]
\centering
\includegraphics[width=\columnwidth]{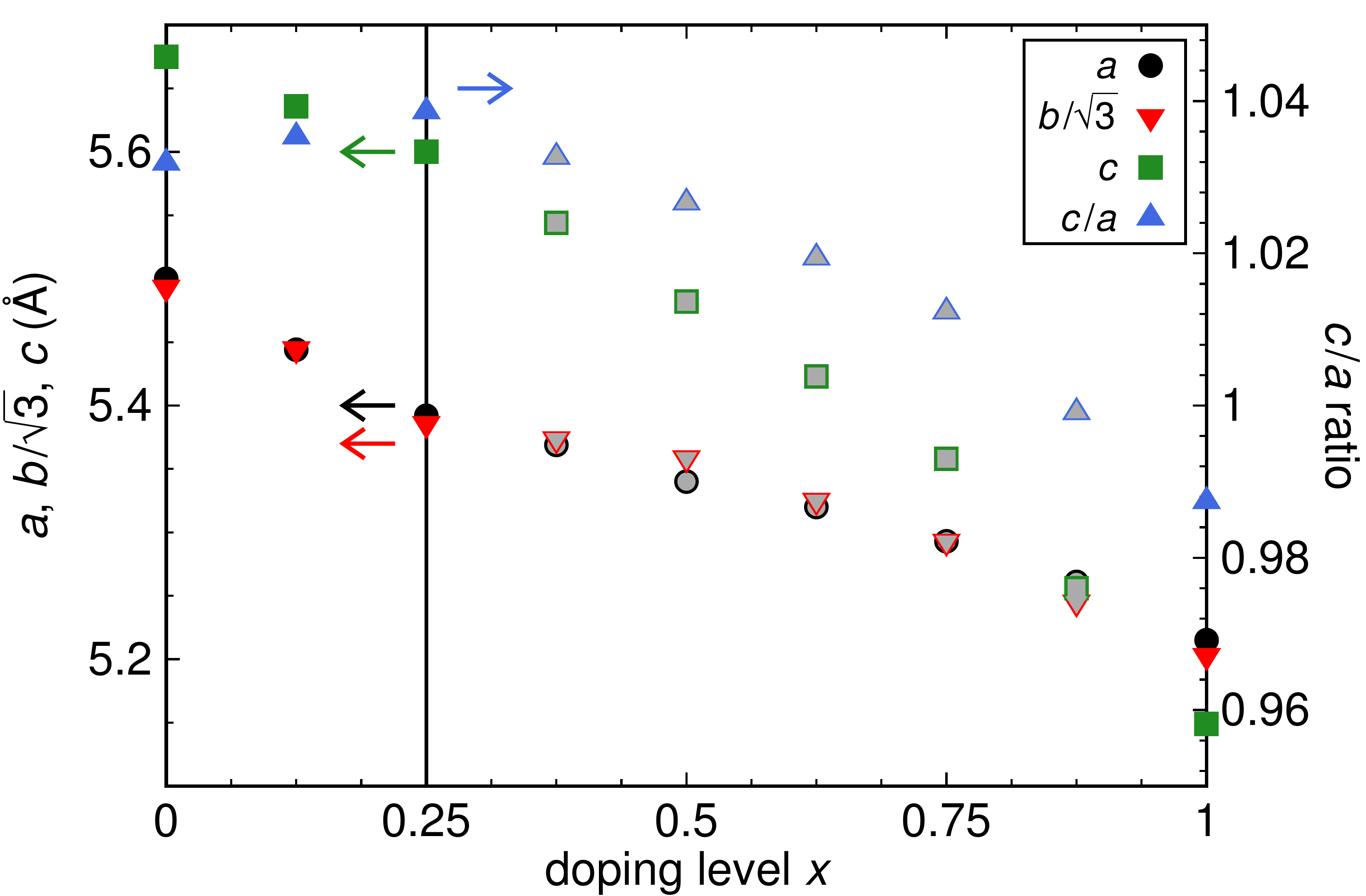}
\caption{(Color online) Calculated lattice parameters of
  (Na$_{1-x}$Li$_{x}$)$_{2}$IrO$_{3}$ within spin-polarized GGA+SOC+U. See the left axis for $a$, $b$,
  and $c$ and the right axis for the $c/a$ ratio.  
   Lattice parameters
 for structures that according to the total energy calculations
  plotted in Fig.~\ref{phase} are only metastable are shown with
  grey symbols only.} \label{latticepar}
\end{figure}

In Fig.~\ref{latticepar} we present the lattice parameters predicted
by spin-polarized GGA+SOC+U calculations ($U=3$~eV, $J=0.5$~eV). In the
range that was accessible experimentally, we find remarkably good
agreement between the calculated lattice parameters and the
experimental values, shown in Fig. \ref{XRD}(b). Although there exists
a small overestimation in the whole range $0\leq x\leq 0.25$, the
trends are caught extremely well and we could even reproduce the
increase in the $c/a$ ratio obtained in the experiment.

\begin{figure}[ht]
\includegraphics[width=\columnwidth]{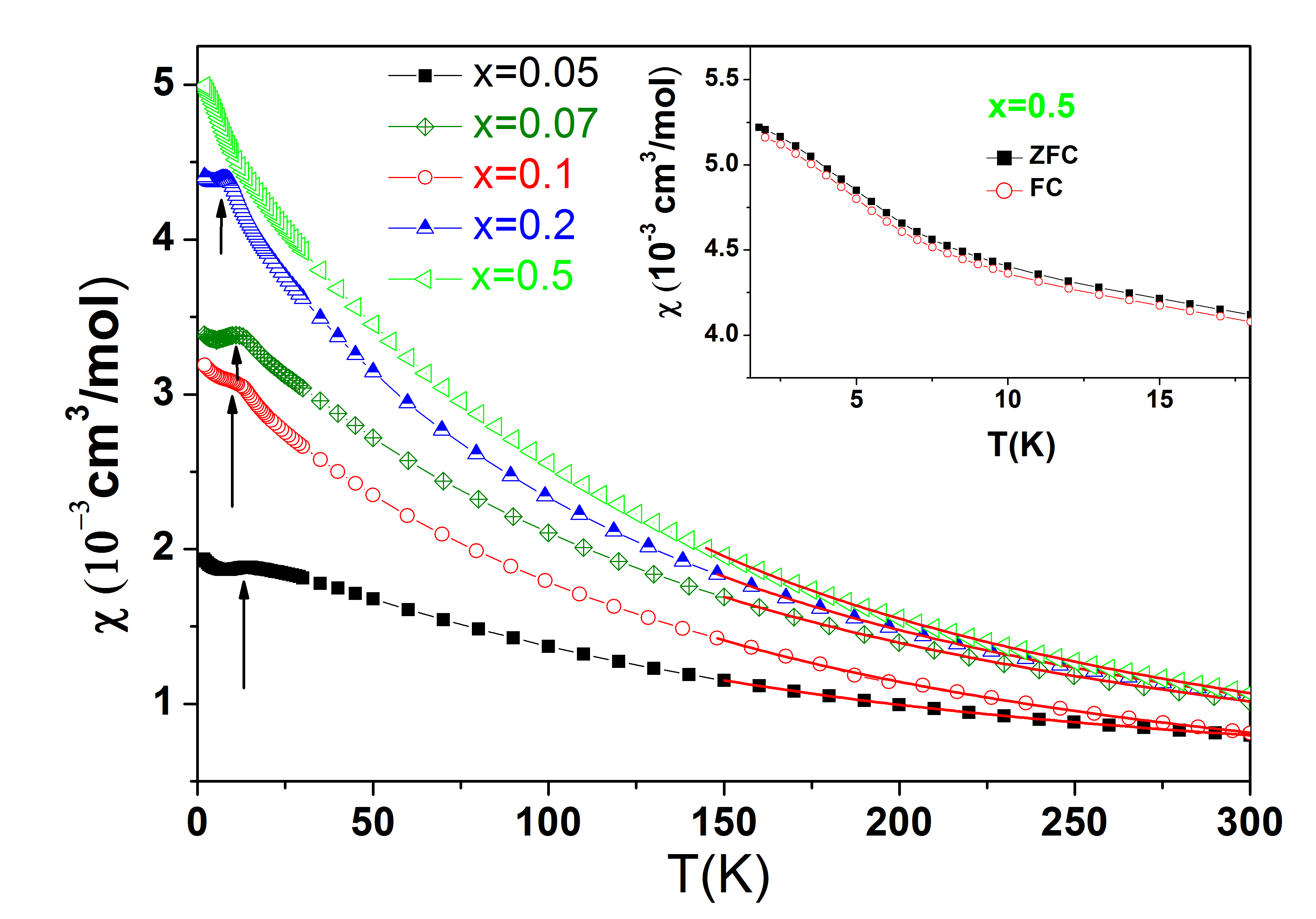}
\caption{(Color online) Magnetic
susceptibility $\chi(T)$ vs T for x= 0.05 to 0.2 and x=0.5. The red line indicates
fitting by CW behavior $\chi= \chi_{0} +\frac{C}{T-\theta_{W}}$. The arrows
mark the positions of $T_{N}$. FC and ZFC measurements for $x=0.5$ are shown in the inset.  }
\label{Mag}
\end{figure}

\textbf{Magnetic Susceptibility:}  In Fig.~\ref{Mag} we show the 
temperature T dependence of the
magnetic susceptibility $\chi(T)=M/H$ for
(Na$_{1-x}$Li$_{x}$)$_{2}$IrO$_{3}$ for dopings $x$ =0.05 to 0.2 measured
at $H=1~T$ between 2 and 300 K. The inverse susceptibility
($\chi^{-1}$) (not shown) and susceptibility ($\chi$) were fitted
to the Curie-Weiss (CW) law $\chi(T) = \chi_{0}
+\frac{C}{T-\theta_{W}}$ (red lines in Fig.~\ref{Mag}) between 150 and
300 K. For all $x$ values measured,  $\chi_{0}\approx10^{-4}$ cm$^{3}$/mol and
$C$= 0.4-0.5 cm$^{3}$ K/mol, while the Weiss temperature
($\theta_{W}$) is dependent on doping (see Fig.~\ref{phase}(b)).
Since for single crystalline Na$_2$IrO$_3$ an anisotropic
susceptibility was observed,~\cite{Singh10} we  expect a certain
anisotropy in the different Li-substituted single crystals as
well. The susceptibility measured on lumps of arbitrary oriented
crystals is therefore different from the average between $\chi_a$
and $\chi_c$ and would not match a perfectly random
polycrystalline sample. This explains a $\approx 20\%$ variation
in the $C$ parameter of the Curie-Weiss fit for the different Li
substituted samples.
$\chi$(T) shows a kink for all measured $x$ (marked
with arrows in Fig.~\ref{Mag}) indicating long range
AF ordering. No spin glass freezing has been observed, as
confirmed by FC-ZFC and ac susceptibility measurements. We
determined the position of maxima  by plotting $\frac{d\chi}{dT}$
vs T where the zero crossing is assigned to the AF transition temperature $T_{N}$. 

\begin{figure}[tbh]
\centering
\includegraphics[width=\columnwidth]{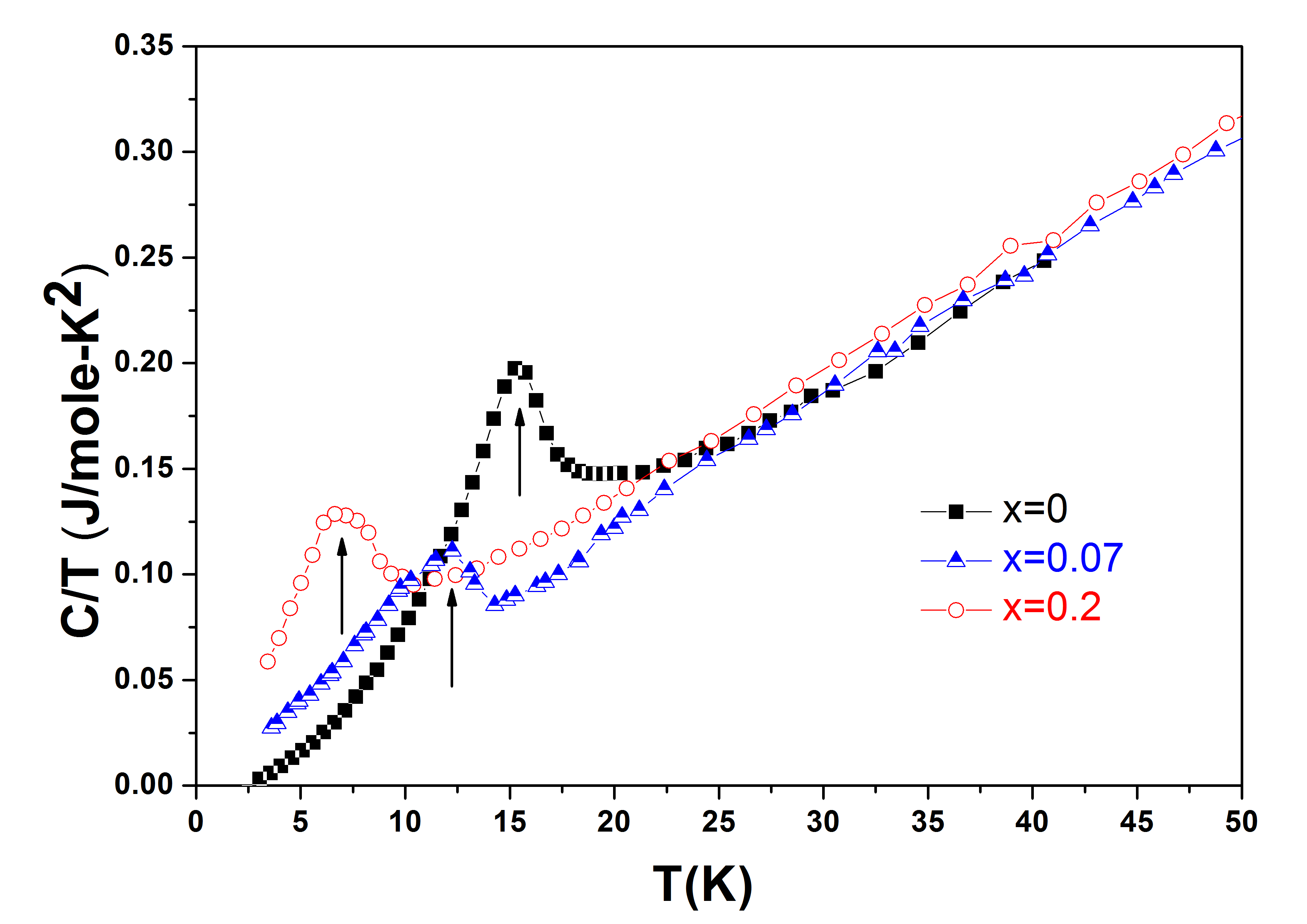}
\caption{(Color online) Heat capacity as $C(T)/T$ of single
phase (Na$_{1-x}$Li$_{x}$)$_{2}$IrO$_{3} $~crystals. The arrows
mark the positions of $T_{N}$. } \label{CbyT}
\end{figure}

\textbf{Heat Capacity:}
Fig.~\ref{CbyT} shows the heat capacity divided by temperature
$(C/T)$ of (Na$_{1-x}$Li$_{x}$)$_{2}$IrO$_{3}$ crystals up to
$x=0.2$. These measurements confirm bulk AF ordering and the extracted
$T_N$ (from the onset of the lambda-like peaks in $C/T$) as a function
of Li doping agrees with the values from the susceptibility
measurements. In order to obtain information on the size of the
ordered moment, we have determined the magnetic entropy from
integration of the magnetic heat capacity ($\Delta C(T)/T$). The
latter was calculated by subtracting the phonon contribution. For
$x=0$ the phonon heat capacity is obtained from the non-magnetic
reference Na$_{2}$SnO$_{3}$ while for $x=0.2$ we use as reference
80$\%$ contribution of Na$_{2}$SnO$_{3}$ and 20$\%$ of
Li$_{2}$SnO$_{3}$. Integration of $\Delta C/T$ vs T reveals values of
the magnetic entropy $\Delta S = 0.2 R\ln2$ and 0.12 R$\ln2$ at T$_N$
for $x=0$ and $x=0.2$, respectively.  This suggests a suppression of
the ordered moment ($0.22~\mu_{B}$ at $x=0$, see
Ref.~\onlinecite{Feng}) by Li substitution, which may be due to
stronger frustration and/or local lattice distortions that affect the
magnetic exchanges.

\subsection{Higher doping ($x>0.25$)}

The systematic suppression of $T_{N}$ with increasing $x$ for
(Na$_{1-x}$Li$_{x}$)$_{2}$IrO$_{3}$~crystals up to $x=0.2$
suggests the possibility of a magnetic quantum phase transition at
larger $x$. However, for larger Li content, i.e., from
$x=0.25$ to $x=0.6$ we see a clear indication of phase separation
in the respective samples. The powder XRD patterns of crushed crystals are shown in Fig.~\ref{XRD}(a,b). Fig.~\ref{XRD}(a) shows that the x=0.6 pattern contains (00n) peaks located close to both pure \liiro (marked by downwards pointing magenta colored arrows) and \nairo (indicated by upwards pointing black colored arrows). A closer inspection of the region near (001) with more different compositions is given in Fig.~\ref{XRD}(b). It shows that for all nominal compositions larger than 0.2 two phases are observed, one close to x=0.2, the other one x=1.  In the single crystal XRD at the higher dopings
$x=0.3,0.4$ the samples showed many  co-existing single
crystal grains compared to the crystals at dopings
 $x$$\leq$0.2 region and the diffraction data could not be consistently
indexed by the same unit cell parameters for all co-existing
grains, suggesting that the samples were not single-phase, but
possibly a mixture of phases with different lattice parameters.

\begin{figure}[ptb]
\centering
\includegraphics[width=\columnwidth]{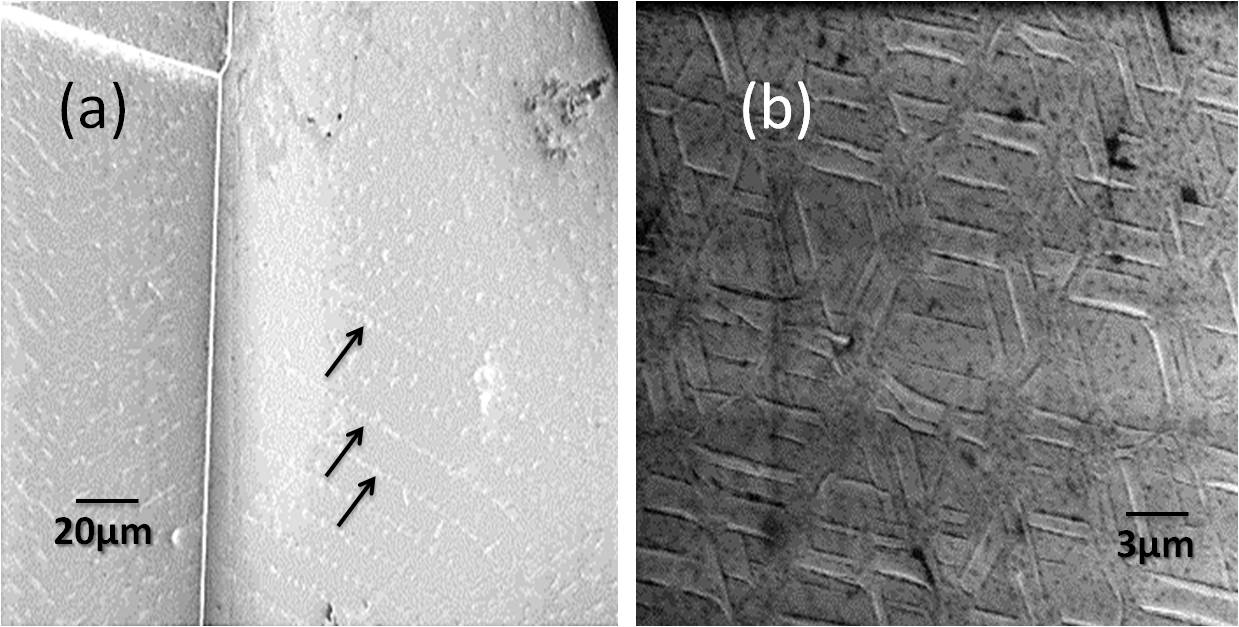}
\caption{SEM picture of (a) $x=0.3$ and (b) $x=0.6$
(Na$_{1-x}$Li$_{x}$)$_{2}$IrO$_{3}$ crystals.} \label{highxp2}
\end{figure}

\begin{figure}[ptb]
\centering
\includegraphics[width=\columnwidth]{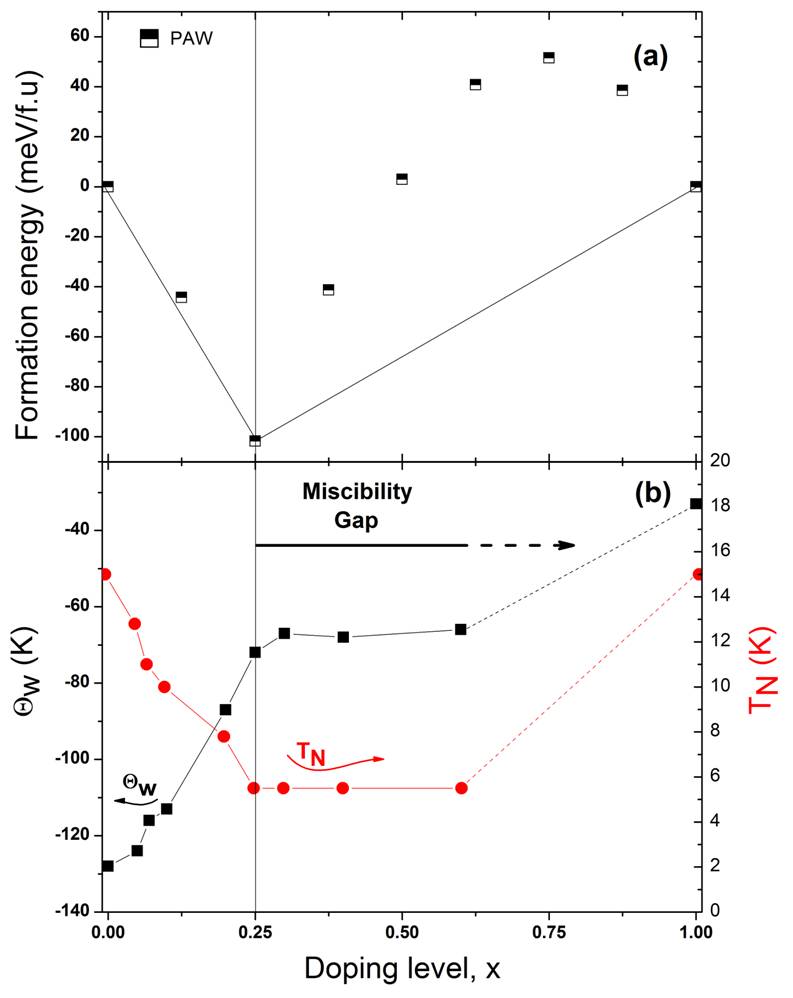}
\caption{(Color online) (a) Phase diagram of \nalio~ obtained from spin-polarized GGA+SOC+U
  total energy calculations. Shown are the formation energies obtained
  with the PAW basis. The vertical line indicates
  the composition at $x=0.25$ which is very stable Na$_3$LiIr$_2$O$_6$
  structure with alternating LiIr$_2$O$_6$ and Na$_3$ layers. (b) Phase
  diagram with $T_{N}$ and CW temperature $\theta_{W}$ of
  (Na$_{1-x}$Li$_{x}$)$_{2}$IrO$_{3}$, data at $x=1$ are from
  Ref.~\onlinecite{Singh12}. The miscibility gap region is indicated
  by the horizontal arrow.}
 \label{phase}
\end{figure}

The two phase scenario is further supported by the results of
scanning electron microscopy (SEM) shown in Fig.~\ref{highxp2} (a) and
(b) for $x=0.3$ and 0.6 crystals, respectively. For $x=0.3$ two phases
were observed. On the lighter contrast lines (marked by arrows in
Fig.~\ref{highxp2}(a)) EDX shows a much lower ratio of Na:Ir (almost
only Ir). Hence this lighter contrast can be attributed to the
Li$_{2}$IrO$_{3}$~phase. For $x=0.6$ hexagonal shaped micro-domains
appear (average size 2-3 $\mu$m).  The SEM picture was taken after
cleaving the crystals and micro-domains of the same size are still
present. EDX measurements show a very small Na:Ir ratio at the domain
boundaries, indicating also Li$_{2}$IrO$_{3}$~micro-domains. In fact
ICPMS indicates (Table~\ref{table:ICPMS}) an increase in Li content
for $x\geq~0.3$, although there is not much change in the lattice
parameters for $x$=0.3 and 0.4 compared to $x$=0.2 (see
Fig.~\ref{XRD}(c)). The trend of change in lattice parameters
significantly deviates after x=0.25. This confirms that in the region
$0.25<x\leq 0.6$ Li is not incorporated into the main
(Na$_{1-x}$Li$_{x}$)$_{2}$IrO$_{3}$~phase but rather forms separate
micro-domains of Li$_{2}$IrO$_{3}$ indicating a miscibility gap in the
phase (see Fig.~\ref{phase}(b)).

\begin{figure}[ptb]
\centering
\includegraphics[width=\columnwidth]{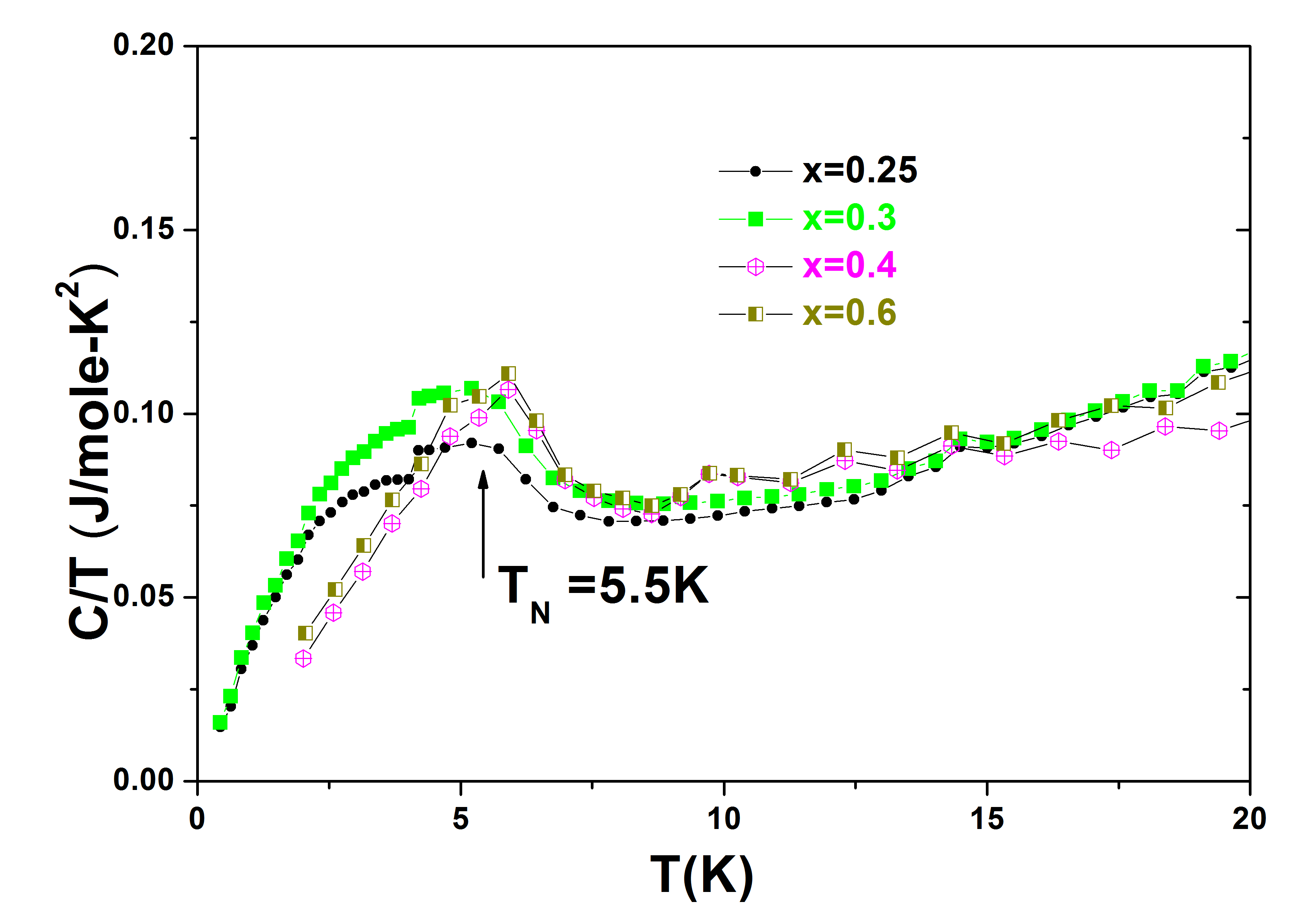}
\caption{(Color online) Heat capacity as $C(T)/T$ of multiphase x$\geq$ 0.25.
(Na$_{1-x}$Li$_{x}$)$_{2}$IrO$_{3} $~crystals. The arrows
mark the positions of $T_{N}$ which is fixed with increasing x.} 
\label{highxp3}
\end{figure}
This is further confirmed when heat capacity is measured for $0.25
\leq x\leq 0.6$. We observe in this whole range a smeared lambda-like
peak at 5.5~K (Fig.~\ref{highxp3}), which implies that $T_N$ does not
depend on doping in this entire range. This means that the magnetic
contribution originates from the main Na$_3$Ir$_2$LiO$_{6}$~phase,
which is not affected by further doping. The micro-domains of
Li$_{2}$IrO$_{3}$ apparently do not exhibit long range order,
presumably due to structural disorder.~\cite{Singh12} For x=0.5
magnetic susceptibility neither shows conventional antiferromagnetic (AF)
 ordering
(Fig.~\ref{Mag}) nor any separation between ZFC-FC
susceptibility(inset) indicative of spin-glass
behavior. We speculate that for this high doping region the presence of a
multidomain \liiro~phase smears out any AF transition in
susceptibility.

Our DFT supercell calculations of (Na$_{1-x}$Li$_{x}$)$_{2}$IrO$_{3}$
at various dopings (see section~\ref{theo}) show that in the
$0<x<0.25$ range, x=0.125 and x=0.25 results are  compatible with a uniform phase within the
computational accuracy. However,
after the Ir$_2$Na planes are completely substituted by Li, further
doping ($x> 0.25$) is energetically unfavorable: for $0.25<x<1$ the
energies of the lowest uniform phases are at least about 30 meV/Ir
higher than those of the separated phases (see
Fig.~\ref{phasewithoutSOC} in Appendix B).
Moreover, the lowest-energy solutions tend to
clusterize on the scale allowed by a given supercell. 
The inclusion of spin-orbit coupling, a Hubbard $U=$~3~eV and magnetism~\cite{SOC+U} leads to an even more pronounced instability towards phase separation ($\gtrsim$~40~meV/Ir), as shown in Fig.~\ref{phase}(a), where the straight line indicates the energy of the corresponding mixture of separated phases.

\section {Conclusions} Based on our structural, thermodynamic, SEM
and magnetic measurements first principles
calculations, we propose the following scenario: in the (Na$_{1-x}
$Li$_{x}$)$_{2}$IrO$_{3}$ system a miscibility gap emerges for
$x>0.25$ (Fig.~\ref{phase}). The stable structure in this region
shows a phase separation into an ordered Na$_3$Ir$_2$LiO$_{6}$
phase, with alternating LiIr$_2$O$_6$ and Na$_3$ planes, and a Li-rich
phase very close in composition to Li$_{2}$IrO$_{3}$. As the
crystal grows, the Na$_3$Ir$_2$LiO$_{6}$ phase nucleates first,
and forms the matrix. We suggest that nucleation for the
Li$_{2}$IrO$_{3}$~phase should start at higher temperature but at the
low temperature it nucleates around multiple centers of the matrix
(Na$_{1-x}$Li$_{x}$)$_{2}$IrO$_{3}$~phase, forming hexagonal
micro-domains.

However, one cannot completely exclude a possible  high
temperature solid-solution phase. One possibility could be that
there may exist a critical temperature of the miscibility gap for
each nominal composition $x\geq 0.25$  above which a metastable single phase
exists and that such temperature is above the crystal growth
 temperature, and therefore it
becomes extremely hard to get single-phase single-crystals in this
doping region. A recent work~\cite{Cao} has claimed single-phase
crystals for $x=0.7-0.9$. Our work reported here shows that in the
doped samples we have synthesized, phase separation occurs for
$0.25\leq x \le 0.6$ and very likely extends also for higher
dopings, so
a detailed investigation of the phase diagram for $0.25< x<1$, both
stable and metastable, is highly desirable. Indeed, the suggested
possibility of a magnetic quantum critical point at $x \sim 0.75$
might be in a metastable region and its nature needs to be
re-examined.

\begin{acknowledgments}
We acknowledge Klaus Simon for ICPMS
measurements, H.S. Jeevan for discussions on crystal growth and
EDX analysis and Yogesh Singh for collaboration. S.M. acknowledges
funding from Erasmus Mundus EURINDIA project. Work in G\"ottingen
has been supported by the Helmholtz Virtual Institute 521 ("New
states of matter and their excitations"). M.A., H.O.J., and R.V.
acknowledge support by the Deutsche Forschungsgemeinschaft through
grant SFB/TR 49 as well as the Centre for Scientific Computing (CSC) in
Frankfurt. Work at Oxford has been supported by EPSRC (UK).
\end{acknowledgments}

\appendix
\section{Single-crystal x-ray refinement results}

In Tables~\ref{tab_NaLi213_5per} to \ref{tab_NaLi213_20per} we list the structures of {\nalio} in the doping range $x=0.05$ to $x=0.20$ as determined by x-ray diffraction.

\begin{table} [b]
\caption{Structural parameters for $x=0.05$ Li-doping from
single-crystal x-ray data at 300~K. ($C2/m$ space group,
$a=5.379(5)$~\AA , $b=9.314(5)$~\AA , $c=5.594(5)$~\AA , $\beta
=108.714(5)^{\circ}$, $Z$=4). $U$ is the isotropic displacement.
The goodness-of-fit(S) was 1.269, $w_{R2}=0.1684$, $R_1=0.0632$
($R_{\rm{int}}=0.0797$, $R_{\sigma}=0.051$).}
\label{tab_NaLi213_5per} 
\begin{tabular}
[c]{lllllll}\hline\hline Atom & Site & $x$ & $y$ & $z$ & Occ &
$U$(\AA $^{2}$)\\ \hline
Ir1 ~ & 4$g$ ~ & 0.5 ~      & 0.1667(1)  & 0 ~~     & 0.849  & 0.0074(6)\\
Na1 ~ & 4$g$ ~ & 0.5 ~      & 0.1667(1)  & 0 ~~     & 0.151  & 0.0074(6)\\
Na2 ~ & 2$a$ ~ & 0 ~        & 0 ~~       & 0 ~~     & 0.498  & 0.0092(8)\\
Ir2 ~ & 2$a$ ~ & 0 ~        & 0 ~~       & 0 ~~     & 0.302  & 0.0092(8)\\
Li2 ~ & 2$a$ ~ & 0 ~        & 0 ~~       & 0 ~~     & 0.2    & 0.0092(8)\\
Na3 ~ & 2$d$ ~ & 0.5 ~      & 0 ~~       & 0.5 ~~   & 1      & 0.021(4)\\
Na4 ~ & 4$h$ ~ & 0.5 ~      & 0.3388(11) & 0.5 ~~   & 1      & 0.019(3)\\
O1 ~  & 8$j$ ~ & 0.758(3) ~ & 0.1732(11) & 0.792(3) & 1 ~~   & 0.013(3)\\
O2 ~  & 4$i$ ~ & 0.720(4) ~ & 0 ~~       & 0.210(4) & 1 ~~   & 0.013(4)\\
\hline\hline
\end{tabular}
\end{table}


\begin{table}[b]
\caption{Same as Table~\ref{tab_NaLi213_5per} for $x=0.10$.
$S=1.467$, $w_{R2}=0.2143$ and $R_1=0.0753$ ($R_{\rm{int}}=0.053$,
$R_{\sigma}=0.0515$).}
\label{tab_NaLi213_10per} 
\begin{tabular}
[c]{lllllll}\hline\hline Atom & Site & $x$ & $y$ & $z$ & Occ & $U$(\AA $^{2}$)\\
\hline
Ir1 ~ & 4$g$ ~ & 0.5       & 0.1668(1)  & 0 ~~     & 0.8303  & 0.0065(4)\\
Na1 ~ & 4$g$ ~ & 0.5       & 0.1668(1)  & 0 ~~     & 0.1697  & 0.0065(4)\\
Na2 ~ & 2$a$ ~ & 0         & 0 ~~       & 0 ~~     & 0.2605  & 0.0171(7)\\
Ir2 ~ & 2$a$ ~ & 0         & 0 ~~       & 0 ~~     & 0.3395  & 0.0171(7)\\
Li2 ~ & 2$a$ ~ & 0         & 0 ~~       & 0 ~~     & 0.4     & 0.0171(7)\\
Na3 ~ & 2$d$ ~ & 0.5       & 0 ~~       & 0.5 ~~   & 1 ~~    & 0.022(3)\\
Na4 ~ & 4$h$ ~ & 0.5       & 0.3384(7)  & 0.5 ~~   & 1 ~~    & 0.023(3)\\
O1 ~  & 8$j$ ~ & 0.757(2)  & 0.1734(8)  & 0.791(2) & 1 ~~    & 0.014(3)\\
O2 ~  & 4$i$ ~ & 0.719(3)  & 0 ~~       & 0.213(3) & 1 ~~    & 0.013(3)\\
\hline\hline
\end{tabular}
\end{table}

\begin{table}
\caption{Same as Table~\ref{tab_NaLi213_5per} for $x=0.15$.
The sample had two twins rotated around the $c^{*}$ axis with
$R_{\rm{int}}=0.168$ and 0.198 for the data sets of reflections,
with the combined goodness of fit values $S=1.778$,
$w_{R2}=0.2679$ and $R_1=0.1151$.} \label{tab_NaLi213_15per} 
\begin{tabular}
[c]{lllllll}\hline\hline Atom & Site & $x$ & $y$ & $z$ & Occ & $U$(\AA $^{2}$)\\
\hline
Ir1 ~ & 4$g$ ~ & 0.5       & 0.1669(2)   & 0 ~     & 0.915  & 0.009(1)\\
Na1 ~ & 4$g$ ~ & 0.5       & 0.1669(2)   & 0 ~     & 0.085  & 0.009(1)\\
Na2 ~ & 2$a$ ~ & 0         & 0 ~~        & 0 ~     & 0.23   & 0.019(3)\\
Ir2 ~ & 2$a$ ~ & 0         & 0 ~~        & 0 ~     & 0.17   & 0.019(3)\\
Li2 ~ & 2$a$ ~ & 0         & 0 ~~        & 0 ~     & 0.6    & 0.019(3)\\
Na3 ~ & 2$d$ ~ & 0.5       & 0 ~~        & 0.5 ~   & 1      & 0.025(6)\\
Na4 ~ & 4$h$ ~ & 0.5       & 0.3394(16)  & 0.5 ~   & 1      & 0.022(4)\\
O1 ~  & 8$j$ ~ & 0.756(4)  & 0.177(2)    & 0.792(4) & 1 ~~  & 0.016(5)\\
O2 ~  & 4$i$ ~ & 0.709(5)  & 0 ~~        & 0.204(5) & 1 ~~  & 0.005(5)\\
\hline\hline
\end{tabular}
\end{table}

\begin{table}
\caption{Same as Table~\ref{tab_NaLi213_5per} for $x=0.20$.
The sample had three twins rotated around the $c^{*}$ axis with
the $R_{\rm{int}}$ parameter between 0.15 and 0.30 for the three
data sets of reflections, with the combined goodness of fit values
$S=2.091$, $w_{\rm{R}2}=0.3019$, $R_1=0.1237$.}
\label{tab_NaLi213_20per} 
\begin{tabular}
[c]{lllllll}\hline\hline Atom & Site & $x$ & $y$ & $z$ & Occ &
$U$(\AA $^{2}$)\\\hline
Ir1 ~ & 4$g$ ~ & 0.5       & 0.1671(1)   & 0 ~~     & 0.9038  & 0.011(1)\\
Na1 ~ & 4$g$ ~ & 0.5       & 0.1671(1)   & 0 ~~     & 0.0962  & 0.011(1)\\
Na2 ~ & 2$a$ ~ & 0         & 0 ~~        & 0 ~~     & 0.0075  & 0.015(2)\\
Ir2 ~ & 2$a$ ~ & 0         & 0 ~~        & 0 ~~     & 0.1925  & 0.015(2)\\
Li2 ~ & 2$a$ ~ & 0         & 0 ~~        & 0 ~~     & 0.8     & 0.015(2)\\
Na3 ~ & 2$d$ ~ & 0.5       & 0 ~~        & 0.5 ~~   & 1       & 0.027(5)\\
Na4 ~ & 4$h$ ~ & 0.5       & 0.3377(14)  & 0.5 ~~   & 1       & 0.030(4)\\
O1 ~  & 8$j$ ~ & 0.759(5)  & 0.1780(17)  & 0.805(5) & 1 ~~    & 0.025(5)\\
O2 ~  & 4$i$ ~ & 0.716(5)  & 0 ~~        & 0.191(5) & 1 ~~    & 0.019(5)\\
\hline\hline
\end{tabular}
\end{table}

\section{Phase diagram obtained from GGA calculations}

\begin{figure}[ptb]
\centering
\includegraphics[width=0.9\columnwidth]{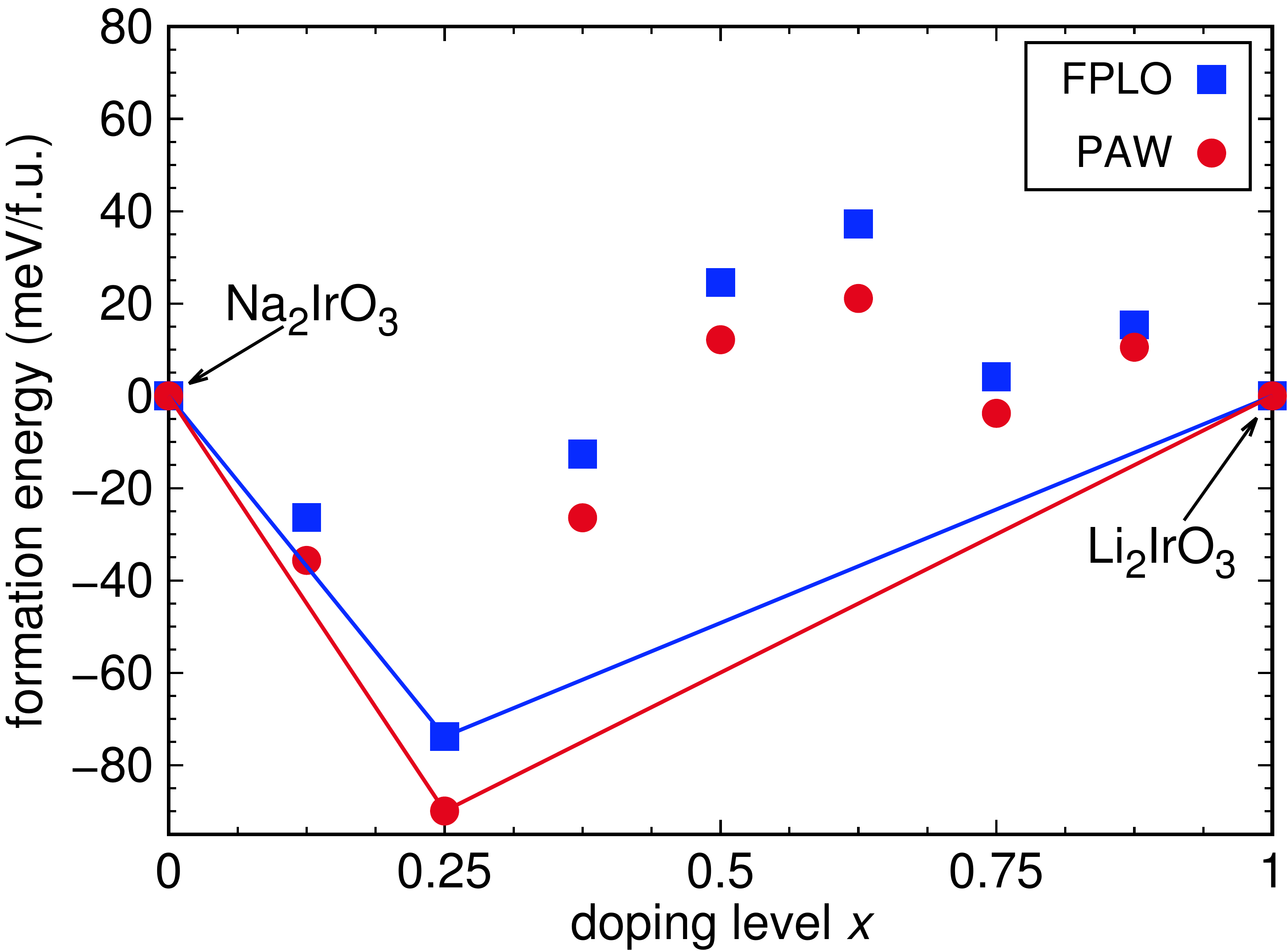}
\caption{(Color online) Phase diagram of \nalio~ obtained from DFT
  total energy calculations.}
 \label{phasewithoutSOC}
\end{figure}

Fig.~\ref{phasewithoutSOC} shows the formation energy of the {\nalio} structures predicted within GGA. The calculations were done with VASP (PAW basis)~\cite{VASP} and with an all electron code (FPLO)~\cite{FPLO}. Qualitatively, the formation energy is very similar to the computationally more expensive spin-polarized GGA+SOC+U results (compare Fig.~\ref{phase}.

\begin{thebibliography}{99}
\bibitem{SOC+U} The structural relaxations under inclusion of SOC and magnetism (stabilized by the use of an on-site Hubbard repulsion of reasonable strength, $U=$~3~eV) were performed for the most stable configurations as obtained in the relaxations described in section III.
\end{thebibliography}
\end{document}